\documentclass[a4paper,10.5pt]{article}

\usepackage[dvips]{graphicx}
\usepackage{float}
\usepackage{epsfig}
\usepackage{graphicx}
\usepackage{longtable}
\usepackage{rotating}
\usepackage{amsmath}
\usepackage{amsfonts}
\usepackage{amssymb}
\usepackage{times}
\usepackage{cite}
\usepackage{dcolumn}
\usepackage{bm}
\usepackage{hyperref}   
\usepackage{multirow}

\usepackage{color}
\usepackage{xspace}
\usepackage{calrsfs}
\DeclareMathAlphabet{\mathpzc}{OT1}{pzc}{m}{it}

\def\ttt#1{\texttt{\small #1}}

\newcommand{\sqrts}{\sqrt{s}}
\newcommand{\sqrtsnn}{\sqrt{s_{_{\mbox{\rm \tiny{NN}}}}}}

\newcommand{\pt}{{p_{\textsc{t}}}}

\newcommand{\pPb}{{\rm{pPb}}}

\newcommand{\mcfm}{{\sc mcfm}}

\newcommand{\toppp}{\ttt{Top++}}
\newcommand{\mt}{m_{t}}

\newcommand{\qqbar}    {q\overline{q}}

\providecommand{\bbbar}{b\overline{b}}
\providecommand{\ttbar}{t\overline{t}}
\newcommand{\alphas}    {\alpha_{\rm s}}

\newcommand*{\eg}{e.g.\@\xspace}
\newcommand*{\ie}{i.e.\@\xspace}

\renewcommand\arraystretch{1.3}

\begin{document}

\begin{center}
{\Large \bf Top quark pair production cross sections at NNLO+NNLL in pPb collisions at $\sqrtsnn$~=~8.16~TeV} \\[0.25cm]
{\large David~d'Enterria$^{1}$ } \\[0.2cm]
{\it $^1$ CERN, EP Department, 1211 Geneva, Switzerland} \\
\end{center}


\begin{abstract}
\noindent
Total and fiducial top pair ($\ttbar$) production cross sections in proton-lead (pPb) collisions
at $\sqrtsnn$~=~8.16 TeV are computed at next-to-next-to-leading-order (NNLO) accuracy including 
next-to-next-to-leading-log (NNLL) gluon resummation, using the CT14 and CT10 proton parton 
distribution functions (PDF), and the EPPS16 and EPS09 nuclear PDF parametrizations for the lead ion. 
The total cross sections amount to $\sigma(\pPb\to\ttbar+X) = 59.0 \pm 5.3${\sc (ct14+epps16)}$\,^{+1.6}_{-2.1}$(scale)~nb,
and 
$57.5 \pm \,^{+4.3}_{-3.3}${\sc (ct10+eps09)}$\,^{+1.5}_{-2.0}$(scale)~nb,
with small modifications with respect to the result computed using the free proton PDF alone.
The normalized ratio of pPb to pp cross sections (nuclear modification factor) 
is $\rm R_{pPb}~=~1.04 \,^{\pm 0.07(EPPS16)}_{\pm0.03(EPS09)}$.
In the lepton+jets decay mode, $\ttbar \to \bbbar\,W(\ell\nu)\,W(\qqbar')$, one expects 600 $\ttbar$ events
in the 180~nb$^{-1}$ integrated luminosity collected in pPb collisions at the LHC so far, after typical acceptance and efficiency losses. 
Differential cross sections at NLO accuracy are presented 
as a function of transverse momentum and rapidity of the top quarks, and of their decay b-jets 
and isolated leptons. 
\end{abstract}

\section{Introduction}
\label{sec:intro}

\begin{figure*}[htbp]
\centering
\includegraphics[width=0.6\textwidth]{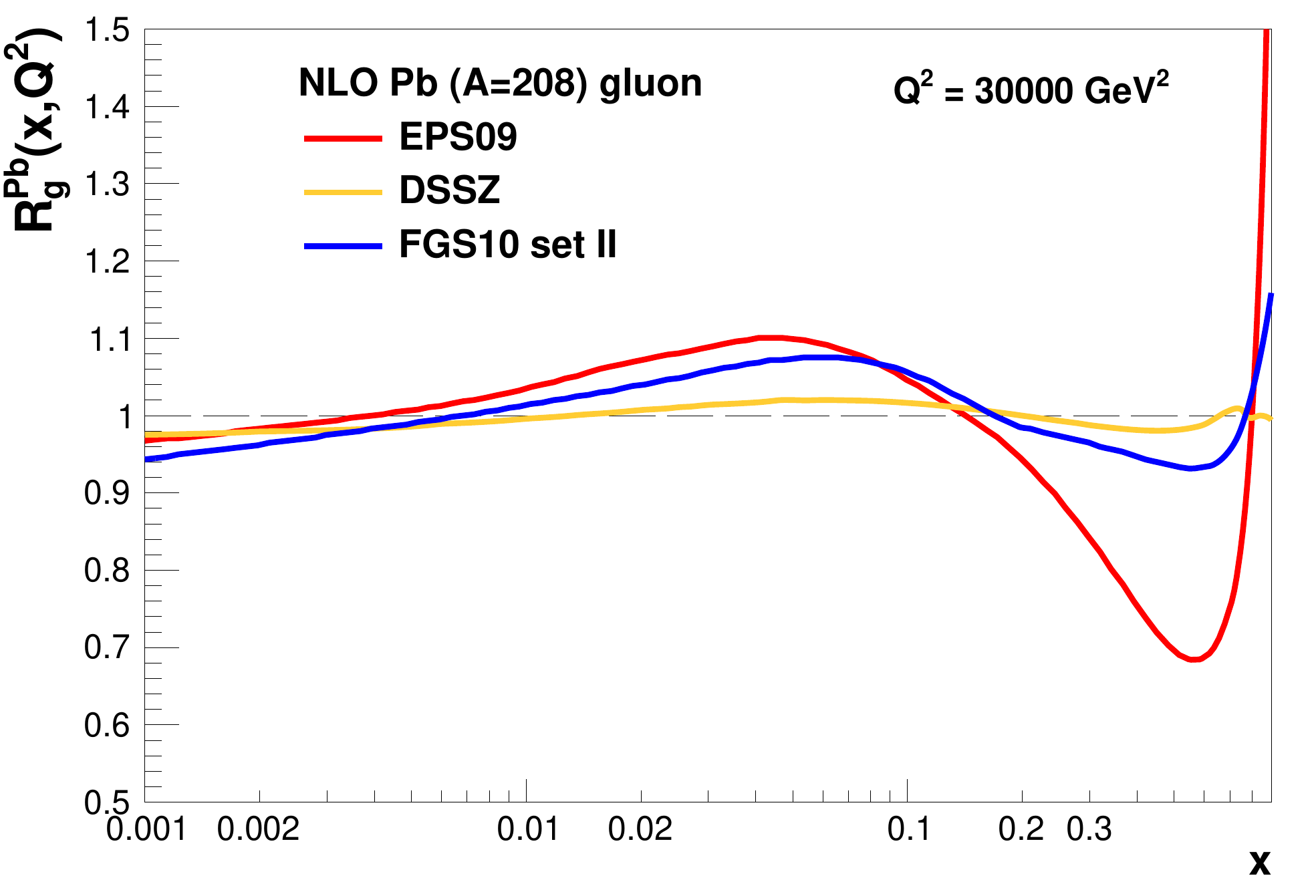}
\caption{Ratio of the lead-to-proton gluon densities in the antishadowing ($x\approx 0.05-0.1$) and EMC 
($x\approx 0.1-0.6$) regions probed by $\ttbar$ production at virtualities $Q^2=\mt^2\approx 3\cdot10^{4}$~GeV$^2$
in pPb collisions at the LHC, for three different NLO nuclear PDF sets: EPS09~\cite{eps09}, DSSZ~\cite{dssz}, and FGS10~\cite{fgs10}.}
\label{fig:nPDFs} 
\end{figure*}

The top quark is the heaviest elementary particle in the Standard Model (SM) and remains unobserved so far in nuclear collisions~\cite{dEnterria:2017jyt}. 
Its cross section in hadronic collisions is dominated\footnote{At NLO, we find that more than 85\% of the $\ttbar$ cross section
at 8.16~TeV involves initial-state gluons from the colliding nucleons.} by pair production in gluon-gluon fusion ($g\,g\to \ttbar+X$), 
which is theoretically computable today with great accuracy via perturbative quantum chromodynamics methods. Calculations 
at next-to-next-to-leading-order (NNLO) including next-to-next-to-leading-log (NNLL) soft-gluon resummations are available 
using \eg\ $\toppp$~\cite{Czakon:2013goa}. Differential $\ttbar$ cross sections are also available at NLO accuracy 
using the \mcfm\ code~\cite{mcfm}. The study of the $\ttbar$ cross section modifications in proton-nucleus compared to pp 
collisions at the same nucleon-nucleon center-of-mass energy ($\sqrtsnn$) provides a novel well-calibrated probe of the 
nuclear gluon density at the LHC~\cite{d'Enterria:2015mgr}, in particular in the unexplored high-$x$ region 
($x\gtrsim 2\,\mt/\sqrtsnn \approx 0.05$) where ``antishadowing'' and ``EMC'' effects are expected 
to modify its shape compared to the free proton case (Fig.~\ref{fig:nPDFs}).

The study of production of top quarks in pPb collisions provides information on the nuclear PDF that is complementary to
that from similar studies with electroweak bosons~\cite{Khachatryan:2015hha,Khachatryan:2015pzs,Aad:2015gta,Alice:2016wka}.
The cross sections of the latter are more sensitive to quark (rather than gluon) densities, at Bjorken-$x$ values about 
twice smaller~\cite{Paukkunen:2010qg}. In addition, a good understanding of top quark in proton-nucleus collisions is crucial 
as a baseline for upcoming studies of heavy-quark energy loss in the quark-gluon-plasma formed in nucleus-nucleus 
collisions~\cite{d'Enterria:2015mgr,Apolinario:2017}.

The top quark decays very rapidly before hadronizing ($\rm \tau_0 = \hbar/\Gamma_{t} \approx 0.15$~fm/c) into 
$t \to W\,b$ with $\sim$100\% branching ratio, with the $W$ themselves decaying either leptonically 
($t\to W(\ell\,\nu)\,b$, 1/3 of the times) or hadronically ($t\to W(\qqbar)\,b$, 2/3 of the times)~\cite{PDG}. 
In Pb-Pb collisions, the charged leptons $\rm \ell = e,\mu$ from the fully-leptonic final-state ($\ttbar\to \bbbar\,2\ell\,2\nu$)
are totally unaffected by final-state interactions, thereby providing the cleanest channel for its observation 
in the complicated heavy-ion environment
~\cite{d'Enterria:2015mgr}, though at the price of a relatively low branching ratio (BR~$\approx$~4\% for the $ee,\,e\mu$ and $\mu\mu$ modes combined). 
In the p-Pb case, thanks to the lower backgrounds and the absence of final-state effects for jets compared to Pb-Pb collisions, 
the leptons+jets final-state ($\ttbar\to \bbbar\,\ell\,\nu\,2j$) is easily measurable and has a much larger branching 
ratio (BR~$\approx$~30\% for $\ell=e,\mu$, increasing to BR$\approx$~34\% when including also $e,\mu$ final-states from $\tau$ decays in 
$\ttbar\to \bbbar\,\tau\,\nu\,2j$)~\cite{PDG} than the purely leptonic decay. In this work, we present predictions for total, 
fiducial, and differential (for the $\ell$+jets channel) cross sections for $\ttbar$ production in p-Pb 
at the center-of-mass energy of the last LHC run for this colliding system ($\sqrtsnn$~=~8.16~TeV). 

\vspace{-0.3cm}
\section{Total and fiducial $\ttbar$ cross sections}
\label{sec:sigma_tot}

The total and differential $\pPb\to\ttbar+X$ cross sections are computed first at NLO accuracy with the \mcfm\ v8.0 code~\cite{mcfm},
using the CT10~\cite{ct10} and CT14~\cite{ct14} proton parton distribution functions (PDF), and the nuclear PDF modification factors 
for the Pb ion given by the EPS09~\cite{eps09} and EPPS16~\cite{epps16} nuclear PDF (nPDF) sets. Then, a 
K-factor, $\rm K=\sigma(NNLO+NNLL)/\sigma(NLO)\approx$~1.20 is computed with $\toppp$ v2.0~\cite{Czakon:2013goa} using the NNLO CT10 and CT14
PDFs alone, in order to scale up the NLO \mcfm\ cross section to NNLO+NNLL accuracy. 
The $\toppp$ and \mcfm\ codes are run with $\rm N_f=5$ flavors, top pole masses set to $\mt=172.5$~GeV, 
default renormalization and factorization scales set to $\mu_{\textsc{r}}=\mu_{\textsc{f}}=\mt$, and QCD coupling set to $\alphas$~=~0.1180. 
All numerical results have been obtained with the latest SM parameters for particle masses, widths and couplings~\cite{PDG}.
The PDF uncertainties include those from the proton and nucleus PDF combined in quadrature, as obtained from the corresponding 
56+40 (52+32) eigenvalues of the CT14+EPPS16 (CT10+EPS09) sets. The theoretical uncertainty linked to the 
scales choices is estimated by modifying $\mu_{\textsc{r}}$ and $\mu_{\textsc{f}}$ within a factor of two with respect to their default
value. In the pp case, such a theoretical NNLO+NNLL setup yields theoretical cross sections in very good agreement with the 
experimental data at $\sqrts$~=~7,~8,~13~TeV at the 
LHC~\cite{Chatrchyan:2011ew,Chatrchyan:2011nb,ATLAS:2012aa,Chatrchyan:2013kff,Aad:2015pga,Khachatryan:2015uqb}. 
The computed nucleon-nucleon cross sections are then scaled by the Pb mass number (A~=~208) to obtain the corresponding pPb cross sections. 




\begin{table*}[htbp!]
\renewcommand\arraystretch{1.4}%
\caption{Total and fiducial (in the $\ell$+jets channel, after typical acceptance cuts) cross sections for 
$\ttbar$ production in pp and pPb collisions at $\sqrtsnn$~=~8.16~TeV at NNLO+NNLL accuracy with different
proton (CT10 and CT14) and ion (EPS09 and EPPS16) PDF. The first and second errors quoted 
  correspond to the PDF and scale uncertainties. 
\label{tab:sigma_ttbar}}\vspace{-0.5cm}
\begin{center}
\begin{tabular}{lccc}\hline
               & \multicolumn{2}{c}{$\sigma(\ttbar)$ total} & $\sigma(\ttbar\to\bbbar\,\ell\nu\,2j)$ fiducial\\\hline
pp  & 265.8 $\,^{+17.4}_{-14.3}${\footnotesize(CT10)}$\,^{+6.9}_{-9.3}$ pb & 272.6 $\,^{+17.2}_{-15.3}${\footnotesize(CT14)}$\,^{+7.0}_{-9.5}$ pb & 31.5 $\,^{+2.0}_{-1.8}${\footnotesize(CT14)}$\,^{+0.8}_{-1.1}$ pb \\
pPb & 57.5 $\,^{+4.3}_{-3.3}${\footnotesize(CT10+EPS09)}$\,^{+1.5}_{-2.0}$ nb & 59.0 $\pm$ 5.3{\footnotesize(CT14+EPPS16)} $\,^{+1.6}_{-2.1}$ nb & 6.82 $\pm$ 0.61{\footnotesize(CT14+EPPS16)} $\,^{+0.18}_{-0.24}$ nb \\\hline
$\rm R_{pPb}$ & 1.04 $\,^{+0.04}_{-0.02}${\footnotesize(EPS09)} & 1.04 $\pm$ 0.07{\footnotesize (EPPS16)} & 1.04 $\pm$ 0.07{\footnotesize (EPPS16)}\\\hline
\end{tabular}
\end{center}
\end{table*}

The total $\ttbar$ cross sections for pp and pPb collisions for various proton and Pb PDF are listed in the first two
columns of Table~\ref{tab:sigma_ttbar}, as well as the nuclear modification factor $\rm R_{\rm pPb} = \sigma_{\rm pPb}/(A\,\sigma_{\rm pp})$.
For pPb, the CT14+EPPS16 calculations give a central $\ttbar$ cross section which is 2.6\% larger than that computed with CT10+EPS09.
The cross section uncertainties linked to the PDF choice 
are $\pm$9\% for CT14+EPPS16, and $+7.5\%/-5.8\%$ for CT10+EPS09. The theoretical $\mu_{\textsc{f}}, \mu_{\textsc{r}}$ scale uncertainties
amount to $+2.5\%/-3.5\%$. Compared to the corresponding pp results, a small net overall antishadowing effect increases 
the total top-quark pair cross section by a meager 4\% for both EPPS16 and EPS09 sets, $\rm R_{pPb}~=~1.04 \,^{\pm 0.07(EPPS16)}_{\pm0.03(EPS09)}$,
where we have considered that proton PDF and theoretical scale uncertainties cancel out in the ratio.

Fiducial top-pair production cross sections can be measured in the $\ell$+jets channel at the LHC taking 
into account their decay branching ratio (BR~$\approx$~30\%), basic ATLAS/CMS detector acceptance 
constraints, 
and standard final-state selection criteria applied to remove $W$+jets and QCD multijet 
backgrounds~\cite{Chatrchyan:2011ew,Chatrchyan:2013kff,Aad:2015pga}, such as:
\begin{itemize}
\item One isolated charged lepton ($\ell = e, \mu$) with $\pt > 30$~GeV, $|\eta|<2.5$, and $R_{\rm isol} = 0.3$
\item Four jets (reconstructed with the anti-$k_{\rm T}$ algorithm with $R=0.5$~\cite{Cacciari:2008gp}) with $\pt > 25$~GeV, $|\eta|<3.0$
\item Lepton-jets separation of $\Delta R (\ell,j) > 0.4$
\end{itemize}
[Often such cuts are sufficient to carry out the $\ttbar$ measurement although, if needed, a threshold 
on the missing transverse momentum from the unobserved $\nu$ can be added.]
The impact of such cuts, evaluated with \mcfm, indicates a 39.5\% acceptance of the total cross section, 
with a very small dependence on the underlying PDF (the maximum difference in acceptances using the proton and ion
PDF amounts to $\pm$0.7\% on the final cross section). The events that pass such selection criteria 
are then often required in addition to have two b-tagged jets. For a typical b-tagging efficiency of 70\%, 
this results in a final combined 
acceptance$\times$efficiency of $\sim$20\% for a $\ttbar$-enriched sample consisting of one isolated charged 
lepton, two light-quark jets, and two b-jets. Taking into account the $\ell$+jets branching ratio, 
the aforementioned acceptance and efficiency, and the 180 nb$^{-1}$ integrated 
luminosities collected by ATLAS and CMS in pPb collisions at 8.16~TeV, we expect about 600 top-quark pair
events reconstructed in this decay channel.

\vspace{-0.3cm}
\section{Differential $\ttbar\to \rm \ell+jets$ distributions}
\label{sec:sigma_diff}

\begin{figure*}[htbp]
\centering
\includegraphics[width=0.49\textwidth]{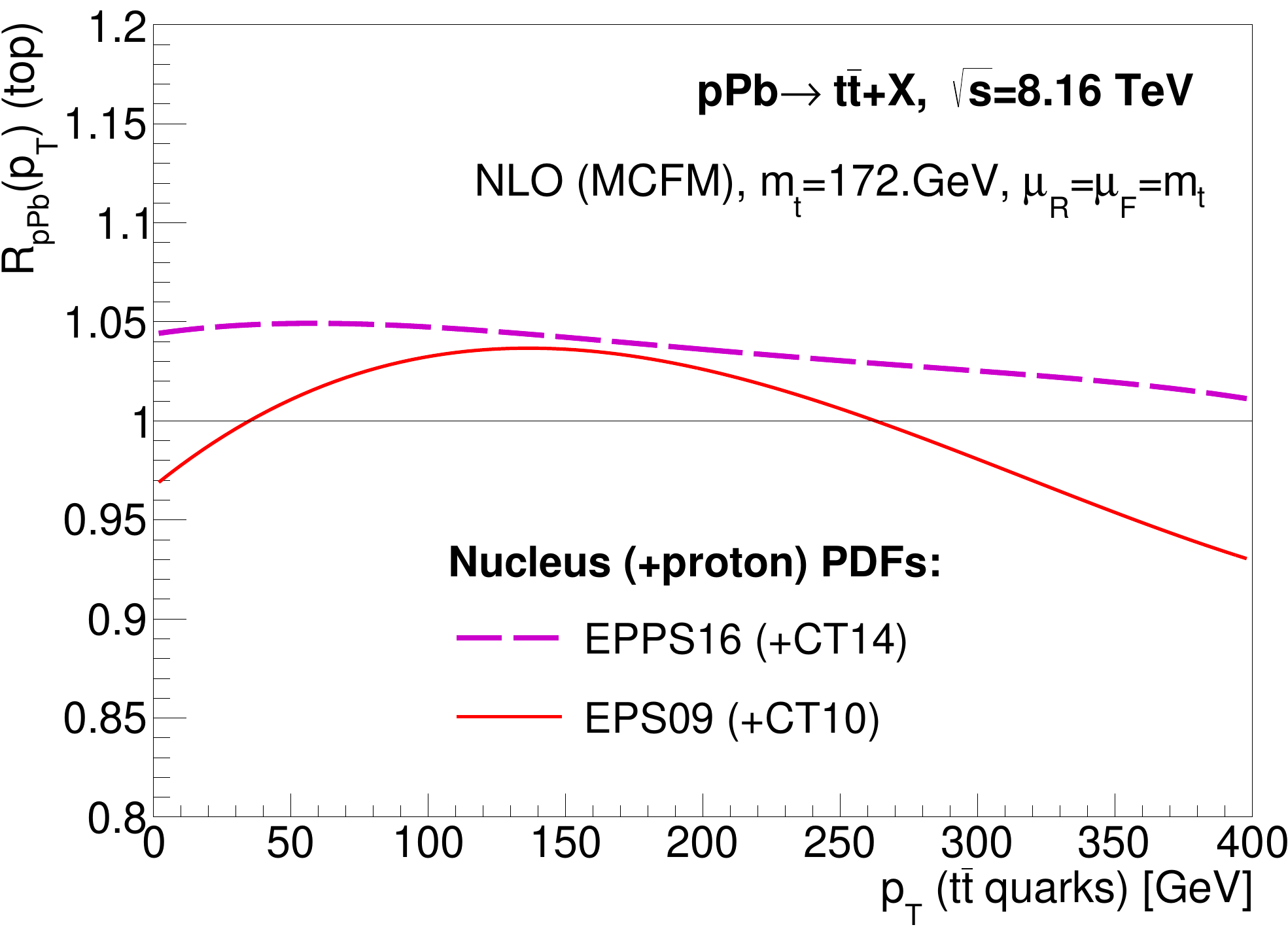}
\includegraphics[width=0.49\textwidth]{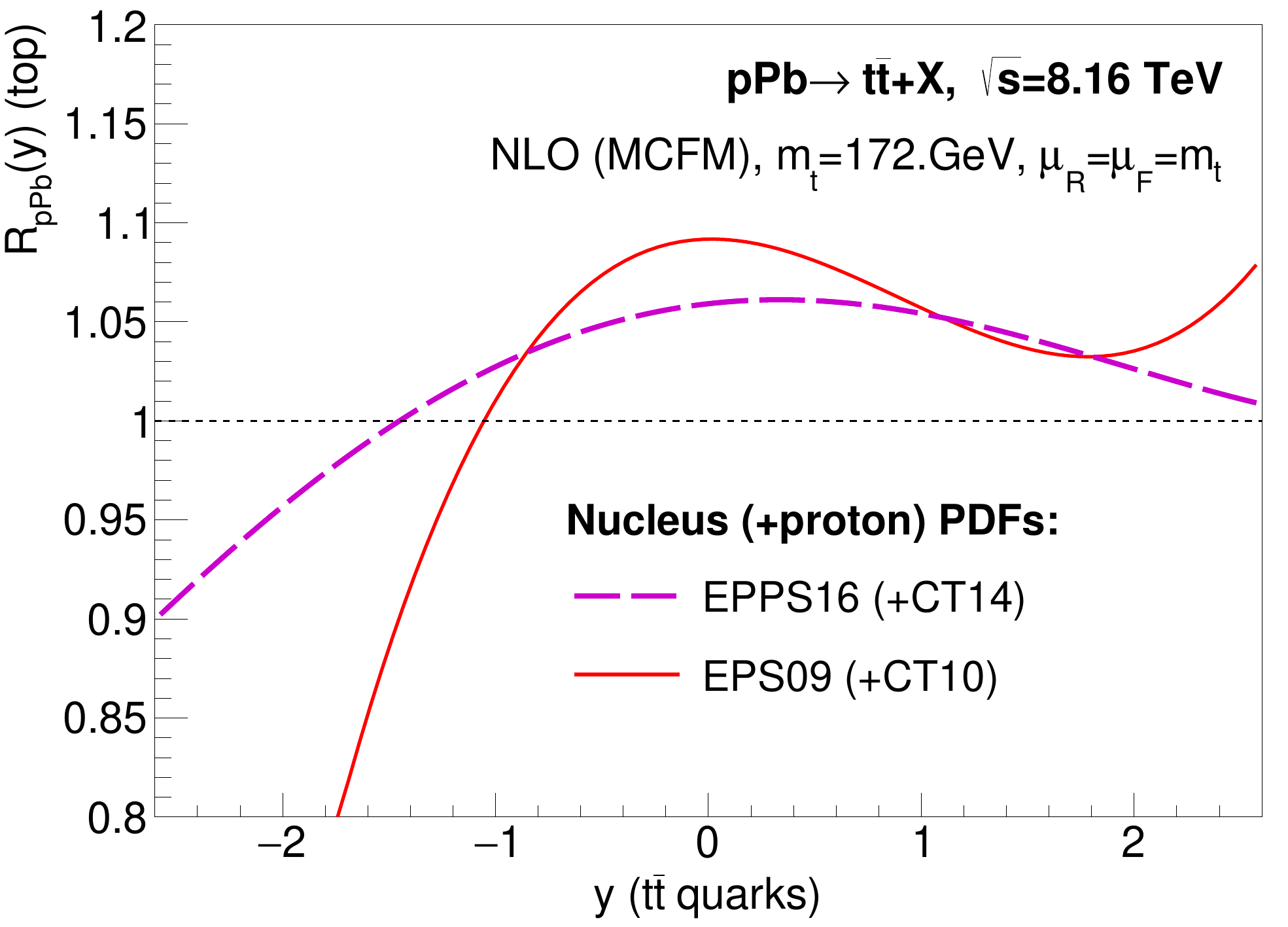}
\includegraphics[width=0.49\textwidth]{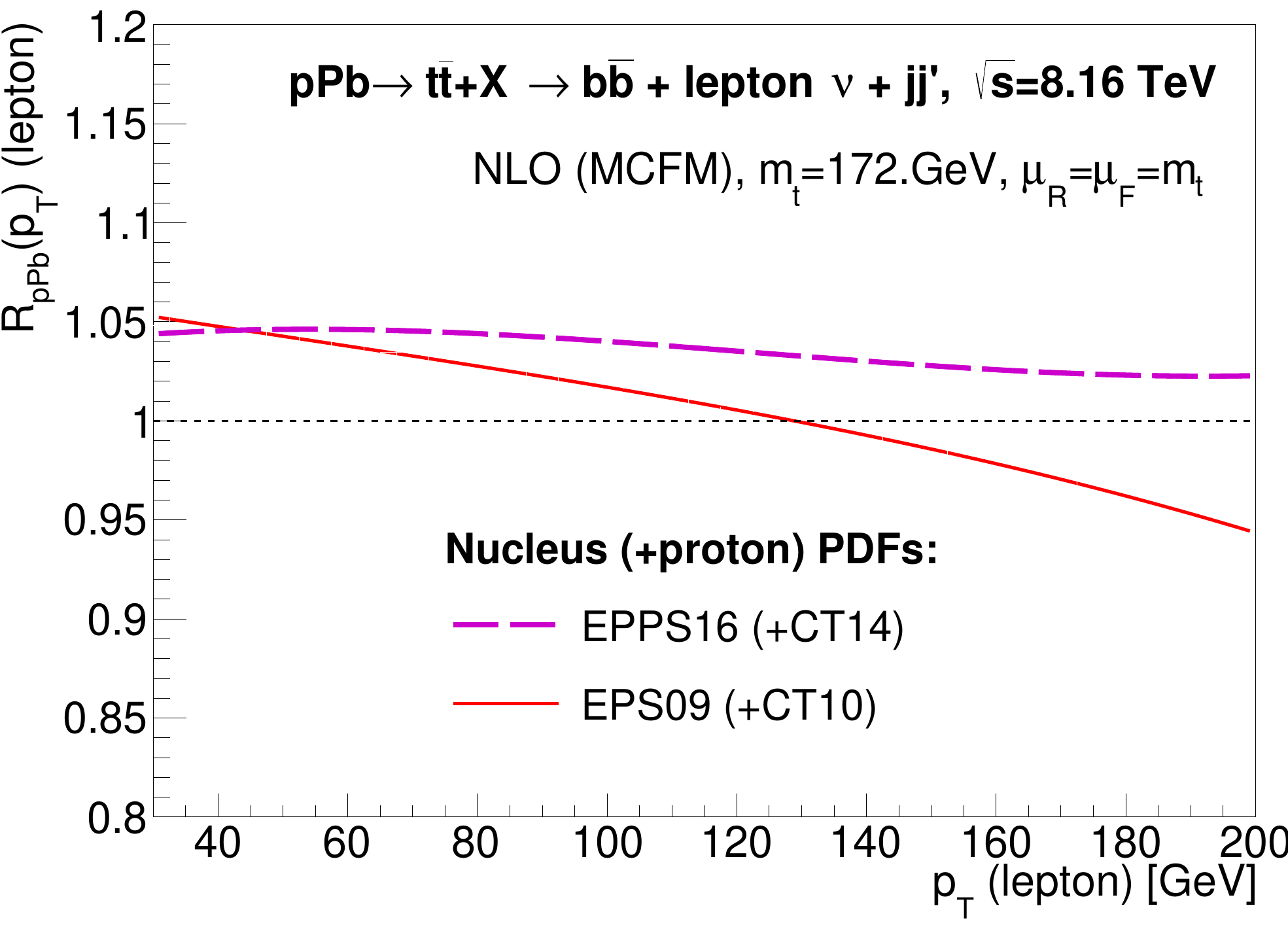}
\includegraphics[width=0.49\textwidth]{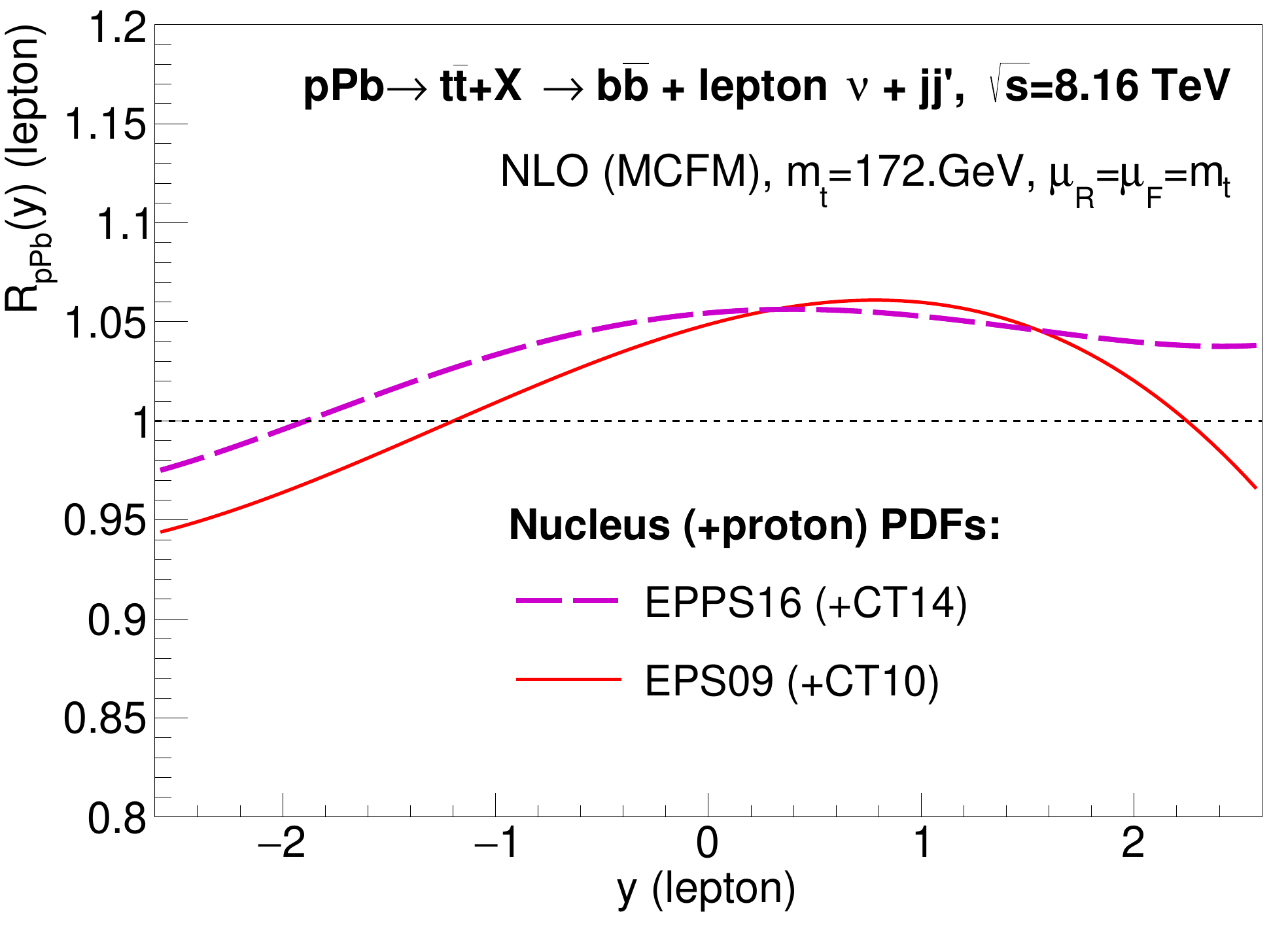}
\includegraphics[width=0.49\textwidth]{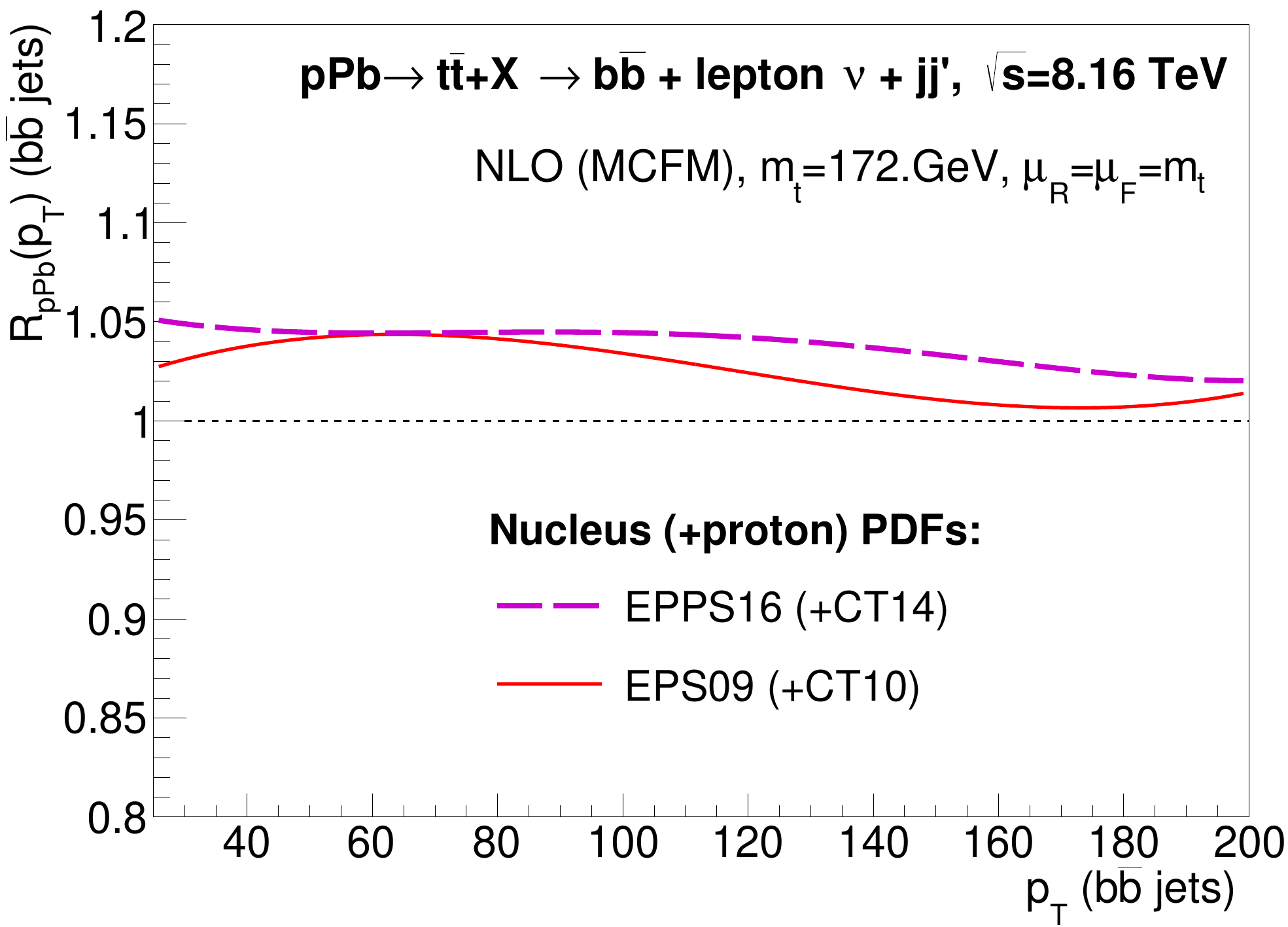}
\includegraphics[width=0.49\textwidth]{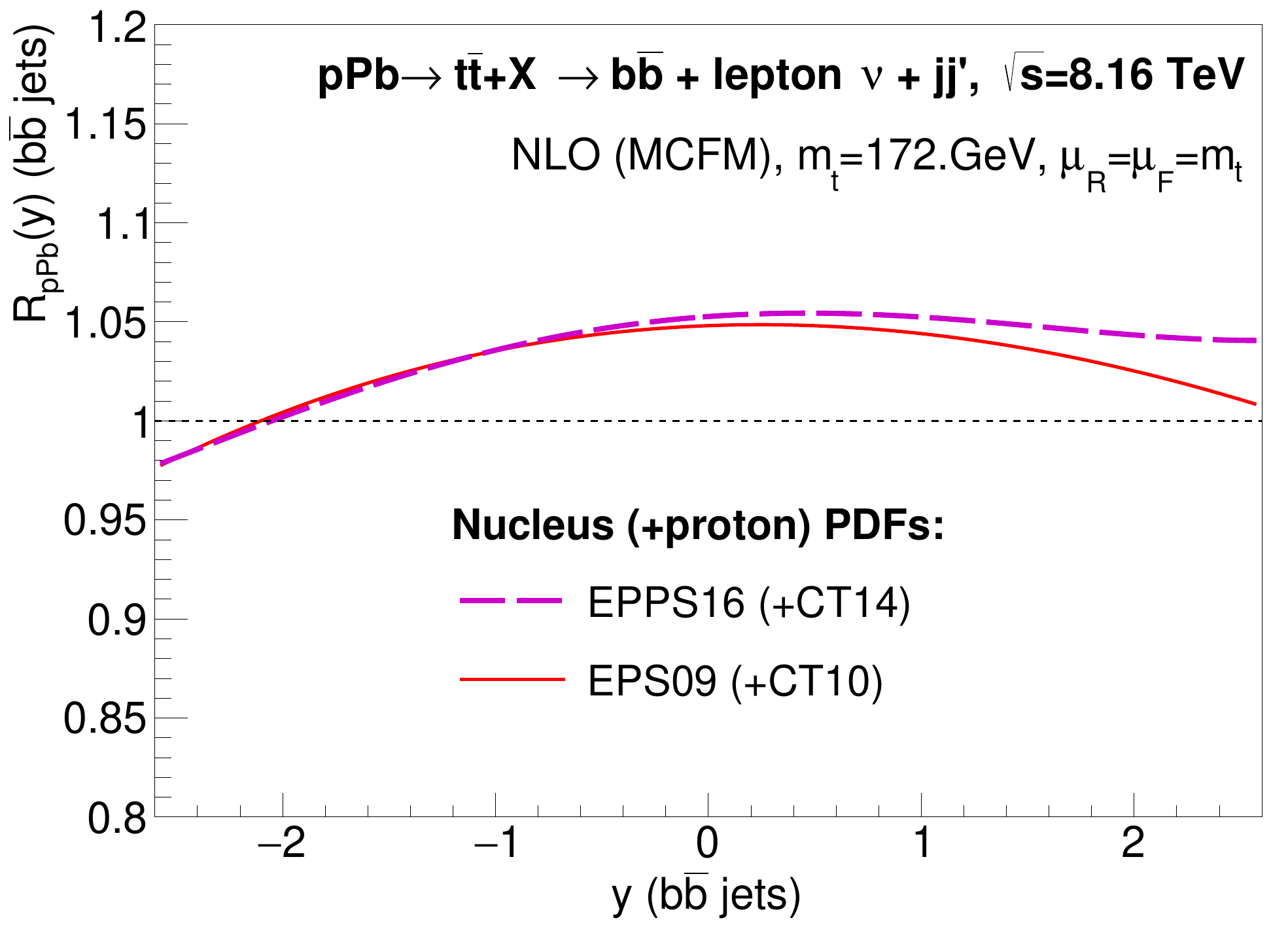}
\caption{Nuclear modification factors as a function of transverse momentum (left) and rapidity (right) 
for $\ttbar$ production in the $\ell$+jets channel at $\sqrtsnn$~=~8.16~TeV for: 
(i) the produced top quarks (top), their (ii) decay isolated leptons (middle), and (iii) 
decay b-jets (bottom), obtained at NLO accuracy with the central PDF sets of CT14+EPPS16 (dashed curves) 
and CT10+EPS09 (solid curves).} \label{fig:RpPb_ttbar} 
\end{figure*}

As seen in the previous section, the total integrated $\ttbar$ cross sections are modified by a few percent only
due to nuclear PDF  effects in pPb compared to pp collisions at 8.16~TeV, $\rm R_{\rm pPb} =$~1.04. However, 
Fig.~\ref{fig:nPDFs} indicates that $gg \to\ttbar$ processes at different $x$ values, \ie\ probed at different 
rapidities and/or transverse momenta of the produced top quarks, should be much more sensitive to the underlying 
positive (antishadowing) and negative (EMC and shadowing) modifications. This was quantitatively confirmed in~\cite{d'Enterria:2015mgr} that showed 
that rapidity distributions of the isolated leptons in the fully-leptonic $\ttbar$ decay mode, are indeed sensitive 
to the underlying nPDF, and can be used to reduce the uncertainties of the EPS09 nuclear gluon density. We present 
a similar study here, but for the $\ell$+jets channel, $\ttbar\to\bbbar\,\ell\nu\,2j$, and using the more updated EPPS16 nPDF set.
Figure~\ref{fig:RpPb_ttbar} shows the nuclear modification ratios, $\rm R_{\rm pPb}(\it X) = (d\sigma_{\rm pPb}/d{\it X})/(A\,d\sigma_{\rm pp}/d{\it X})$ 
as a function of  transverse momentum ($X=\pt$, left panels) and rapidity ($X=y$, right panels)
for (i) the produced top quarks (top), (ii) the decay isolated-leptons (middle), and 
(iii) the decay b-jets (bottom) 
as obtained using the EPPS16 (dashed curves) and EPS09 (solid curves). We note that any effect related 
to the choice of the proton PDF (CT10 or CT14) mostly cancels out in the pPb/pp ratio, which 
is mostly sensitive to modifications of the nuclear gluon densities alone. 
The effect of antishadowing (shadowing or EMC) in the nPDF results in small 5--10\% enhancements (deficits) 
in the distributions at  lower (higher) $\pt$ values as well as at central (forward and backward) rapidities 
$y\approx$~0 ($|y|\gtrsim$~2). In general, the effects are larger for the originally produced 
top quarks than for their decay products (isolated leptons and b-jets), but nonetheless visible also for the
latter.

\vspace{-0.3cm}
\section{Summary}
\label{sec:summary}

Total, fiducial, and differential cross sections for top-quark pair production in proton-lead
collisions at $\sqrtsnn$~=~8.16~TeV have been computed at up to NNLO+NNLL accuracy using
the CT14 and CT10 proton PDF and the EPPS16 and EPS09 nuclear PDF parametrizations. The total cross sections amount to 
$\sigma(\pPb\to\ttbar+X) = 59.0 \pm 5.3${\sc (ct14+epps16)}$\,^{+1.6}_{-2.1}$(scale)~nb,
and $57.5 \pm \,^{+4.3}_{-3.3}${\sc (ct10+eps09)}$\,^{+1.5}_{-2.0}$(scale)~nb,
with few percent modifications with respect to the result obtained using the free proton PDF alone,
$\rm R_{pPb}~=~1.04 \,^{\pm 0.07(EPPS16)}_{\pm0.03(EPS09)}$. In the lepton+jets decay mode, 
$\ttbar \to \bbbar\,W(\ell\nu)\,W(\qqbar')$, one expects about 600 $\ttbar$ events in the 180~nb$^{-1}$ 
integrated luminosity collected at the LHC, after typical ATLAS/CMS acceptance and efficiency losses. 
Ratios of $\ttbar$ differential cross sections in pPb over pp collisions as a function of 
the transverse momentum and rapidity of the charged decay leptons and of the b-jets are sensitive to
the size of antishadowing and EMC gluon density modifications at high virtualities
in the nucleus. Precise differential measurements of top-quark pair production provide thereby 
a novel tool to study the nuclear parton distribution functions in a so-far unexplored kinematical
regime.\\

\vspace{-0.3cm}
\noindent {\bf Acknowledgments--} Discussions with Hannu~Paukkunen on the interface of the EPPS16 PDF 
parametrization with \mcfm\ are gratefully acknowledged.


\end{document}